\documentclass[aps,prb,twocolumn,superscriptaddress,amsmath,amssymb,notitlepage]{revtex4-2}
\usepackage{graphicx}
\usepackage{amsmath}
\usepackage{amssymb}
\usepackage{float}
\usepackage{color}

\usepackage{cancel,xcolor}
\usepackage{indentfirst}
\usepackage{CJKulem}
\usepackage{soul,xcolor}
\setstcolor{red}

\usepackage{ulem}
\usepackage{cancel}
\usepackage[pdfpagemode=UseNone,pdfstartview=FitH,colorlinks=true,linkcolor=blue,urlcolor=blue,anchorcolor=blue,citecolor=blue]{hyperref}
\definecolor{cyan}{rgb}{0.0,1.0,1.0}

\newcommand{\beq}{\begin{equation}}
\newcommand{\eeq}{\end{equation}}
\newcommand{\beqa}{\begin{eqnarray}}
\newcommand{\eeqa}{\end{eqnarray}}

\def\Aa2#1{\textcolor{magenta}{#1}}

\def\Aa1#1{\textcolor{blue}{#1}}

\def\prb#1{{ Phys.\ Rev. B\/} {\bf#1}}

\begin{document}

\title{The puzzle of bicriticality in the XXZ antiferromagnet}

\author{Amnon Aharony}
\email{aaharonyaa@gmail.com}
\affiliation{ School of Physics and Astronomy, Tel Aviv University, Tel Aviv 6997801, Israel}

\author{Ora Entin-Wohlman}
\email{orawohlman@gmail.com}
\affiliation{ School of Physics and Astronomy, Tel Aviv University, Tel Aviv 6997801, Israel}

\begin{abstract}
Renormalization-group  theory predicts that the XXZ antiferromagnet in a magnetic field along the easy Z-axis  has asymptotically either a tetracritical  phase-diagram  or a triple point  in the field-temperature plane.  Neither experiments nor Monte Carlo simulations procure such phase diagrams. Instead, they find a bicritical phase-diagram. Here this discrepancy is resolved:
after generalizing a ubiquitous condition identifying the tetracritical point, we employ  new renormalization-group recursion relations near the isotropic fixed point,  exploiting group-theoretical considerations and using accurate exponents at three dimensions.  These show that the experiments and simulations  results  can only be understood if their trajectories  flow towards the fluctuation-driven first order transition (and the associated triple point), but reach this limit only for prohibitively large system sizes or correlation lengths. In the crossover region one expects a bicritical phase diagram, as  indeed is observed.
A similar scenario may explain puzzling discrepancies between simulations and renormalization-group predictions for a variety of other  phase diagrams with competing order parameters.

\end{abstract}


\date{\today}
\maketitle

\noindent{\bf Introduction.} Natural systems show behaviors ascribed to
fluctuations on many length scales (e.g., critical phenomena, fully-developed
turbulence, quantum field-theory, the Kondo effect,
and  polymers  described by self-avoiding walks). These behaviors can be treated by the renormalization
group (RG)  theory~\cite{wilson,RG,DG}: gradually eliminating short-range details, during which the system size $L$ and the correlation length $\xi$ rescale to $L\rightarrow L(\ell)=L/e^\ell$ and $\xi\rightarrow \xi(\ell)=\xi/e^\ell$ ($\ell$ is the number of RG iterations), the parameters characterizing the
 system can `flow' to a `stable' fixed point (FP), which determines universal power-laws describing physical quantities.
 Varying the parameters can lead to an instability of a FP (with one or more parameters becoming 'relevant' and 'flowing' away from it, as $e^{\lambda \ell}$, with a positive `stability exponent' $\lambda$), generating  transitions between different universality classes.
Although in most cases the predictions of the RG have been confirmed experimentally and/or by numerical simulations, some puzzling discrepancies still await explanations. Here we resolve one such puzzle, involving the phase transitions between competing ordered phases. As listed  e.g. in Refs. \onlinecite{KNF} and \onlinecite{AEK}, phase diagrams with competing order parameters arise in a variety of physical examples. Some of these are mentioned below, after analyzing  the phase diagram of the anisotropic antiferromagnet in a magnetic field.

\begin{figure*}[htb]
\vspace{-2.6cm}
\includegraphics[width=1.\textwidth]{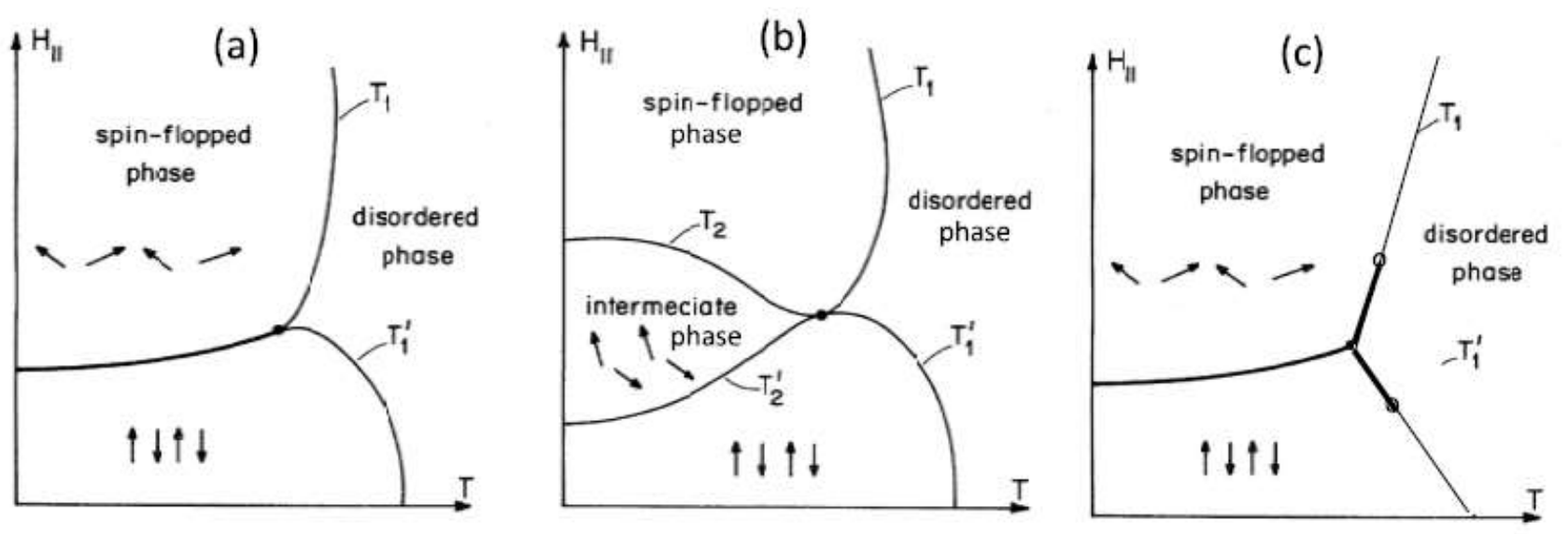}
\vspace{-18.6cm}
\caption{  Possible phase-diagrams for the XXZ antiferromagnet in a longitudinal magnetic field. (a) Bicritical phase diagram.  (b) Tetracritical phase diagram. (c) Diagram with a triple point.  Thick lines - first-order transitions. Thin lines - second-order transitions. The first-order transition lines between the ordered phases and the disordered paramagnetic phase end at tricritical points (small empty circles). After Refs. \onlinecite{bruce,mukamel}.  $T^{}_1$ and $T'^{}_1$ are the transition lines between the ordered phases and the paramagnetic phase. $T^{}_2$ and $T'^{}_2$ are the lines second-order lines which border the mixed phase.}
\label{1}
\end{figure*}

A uniaxially anisotropic XXZ antiferromagnet has long-range  order (staggered magnetization)  along its easy axis, Z. A magnetic field $H^{}_\parallel$ along that axis causes a spin-flop transition into a phase with order in the transverse plane, plus a small ferromagnetic order along Z. Experiments~\cite{king,Shapira} and Monte Carlo simulations on three-dimensional lattices ~\cite{selke0,selke,landau} typically find a {\it bicritical} phase diagram in the temperature-field $T-H^{}_\parallel$ plane [Fig. \ref{1}(a)]: a first-order transition line between the two ordered phases, and two second-order lines between these phases and the disordered (paramagnetic) phase, all meeting at a {\it bicritical point}.
 Recently, the spin-flop transition in XXZ antiferromagnets has raised renewed interest~\cite{seebeck}, related to possible spintronic applications of the Seebeck effect near that transition. Simulations in that paper also seem to find a bicritical phase diagram.

\vspace{.4cm}
\noindent{\bf  History.} The early RG calculations~\cite{KNF} were based on low-order expansions in $\epsilon=4-d$, where $d$ is the spatial dimensionality. These calculations found that the (rotationally-invariant){\it isotropic} FP is stable at $d=3$, yielding asymptotically the bicritical phase diagram.  These calculations also found that the isotropic FP  becomes unstable as the total number of spin components $n$ ($=3$ in our case) increases beyond a threshold $n^{}_c(d)$, and estimated that $n^{}_c(3)>3$. For $n>n^{}_c(d)$ they found a stable {\it biconical} FP.  Had the RG trajectories flown
 to that FP, the first-order line between the two ordered phases would be replaced by an intermediate (mixed) phase, bounded by two second-order lines, and all four second-order lines  would have met at a {\it tetracritical point} [Fig. \ref{1}(b)]~\cite{KNF,bruce,aaDG}.  In addition, if the system parameters are initially outside the region of attraction of that PF, the bicritical point turns into a {\it triple point}, and the transitions between the ordered phases and the disordered paramagnetic phase become first-order near that point, turning second-order only at finite distances from it [Fig. \ref{1}(c)]~\cite{mukamel}.

 However,  the  $\epsilon-$expansions diverge, and low-order calculations are not reliable~\cite{brezin1}.  One way to overcome this divergence is to use resummation techniques, e.g.,  by taking into account the singularities of the series' Borel transforms~\cite{vicari},  and extrapolating the results to $\epsilon=1$. These yielded three stability exponents for the isotropic FP, $\lambda^{}_{0,2,4}$.
   The small exponent $\lambda^{}_4$ also describes the (in)stability against a cubic perturbation~\cite{AAc,aaDG}, and it vanishes at $n=n^{}_c(d)$.  The same resummation techniques  (carried out on sixth-order $\epsilon-$expansions) have been applied to the latter problem~\cite{eps6}. The results were compared with  a resummation of the sixth-order perturbative (divergent) expansions in the original field-theory coefficients at $d=3$~\cite{6loops}, with recent bootstrap calculations~\cite{boot}, with Monte Carlo simulations~\cite{hasen} and with high-temperature series (for $\lambda^{}_0$)~\cite{butera}. An updated table of these results  appears in Ref. \onlinecite{boot}. The agreement between all the techniques indicates the accuracy of the exponents:
 \begin{align}
 \lambda^{}_0\approx -0.78,\ \ \lambda^{}_2\approx -0.55,\ \ \lambda^{}_4\approx 0.01.
 \label{exps}
 \end{align}
 Since $\lambda^{}_4>0$, the isotropic fixed point is unstable at $d=3$, and $n^{}_c(3)<3$, contradicting previous estimates \cite{KNF,aaDG}. Therefore, as explained below, the bicritical phase diagram should be replaced by the tetracritical or the triple one, but neither of these agrees with the experiments or the simulations.

The field theoretical analysis is based on the Ginzburg-Landau-Wilson (GLW) Hamiltonian density~\cite{KNF},
\begin{align}
{\cal H}({\bf r})=&\big(|{\boldmath{\nabla}}{\bf S}|^2+t|{\bf S}|^2\big)/2+U^{}_2+U^{}_4,\\
U^{}_2&=g\big[ |S^{}_\parallel|^2-|{\bf S}|^2/3\big], \\
U^{}_4&=u^{}_\parallel|S^{}_\parallel|^4+ u^{}_\perp |{\bf S}^{}_\perp|^4+2 u^{}_\times |S^{}_\parallel|^2|{\bf S}^{}_\perp|^2,
\label{U4}
\end{align}
with the local three-component ($n=3$) staggered magnetization,  ${\bf S}({\bf r})\equiv \big(S^{}_\parallel({\bf r}),{\bf S}^{}_\perp({\bf r})\big)$.
For $g=0$ and $u^{}_\parallel=u^{}_\perp=u^{}_\times=u$, ${\cal H}$ reduces to the isotropic Wilson-Fisher Hamiltonian~\cite{wilson,RG,DG}, which has an (isotropic) FP at $u=u^I$.~\cite{commu}

\vspace{.4cm}
\noindent{\bf  Group theory.}
 A priori, at $g=0$, the stability of the isotropic FP against  symmetry-breaking perturbations requires an analysis of 15 terms in the GLW Hamiltonian, which are quartic in the spin components, $S^{}_\alpha S^{}_\beta S^{}_\gamma S^{}_\delta$. Group-theoretical arguments showed that these terms split into subsets of $1+5+9$ terms, and all the terms within a subgroup have the same stability exponent, listed in Eq. (\ref{exps})~\cite{wegner,zan,vicari,hasen,vicrev}.  In our case, [$O(3)\Rightarrow O(1)\bigoplus O(2)$],
 the three exponents are associated with the following combinations of quartic terms:
\begin{align}
{\cal P}^{}_{4,0}&\equiv |{\bf S}|^4,\ \ \ {\cal P}^{}_{4,2}\equiv |{\bf S}|^4[x-1/3],\nonumber\\
{\cal P}^{}_{4,4}&\equiv |{\bf S}|^4\big[x(1-x)-(1+x)/7+2/35\big],
\end{align}
where $x=S^{2}_\parallel/|{\bf S}|^2$.
 The largest (negative) exponent $\lambda^{}_0$ corresponds to the stability within the $O(3)-$symmetric case, ${\cal P}^{}_{4,0}$. In our case, the exponent $\lambda^{}_2$ corresponds the a term which splits the $O(3)$ isotropic symmetry group  into $O(1)\bigoplus O(2)$.
Similar to $U^{}_2$, ${\cal P}^{}_{4,2}$ `prefers' ordering of $S^{}_\parallel$ or of ${\bf S}^{}_\perp$. The smallest exponent $\lambda^{}_4$ describes the crossovers away from the isotropic FP, towards either the biconical or the cubic FP.
Writing the quartic terms as
\begin{align}
U^{}_4=(u^I+p^{}_0) {\cal P}^{}_{4,0}+p^{}_2{\cal P}^{}_{4,2}-p^{}_4{\cal P}^{}_{4,4},
\label{U4n}
\end{align}
with arbitrary coefficients $p^{}_i,~i=0,2,4$ (which vanish at the isotropic FP), implies the linear recursion relations near the isotropic FP,
\begin{align}
 dp^{}_i/d\ell\approx\lambda^{}_ip^{}_i\ \ \  \ \Rightarrow\ \ \ p^{}_i(\ell)=p^{}_i(0)e^{\lambda^{}_i\ell}.
 \label{pil}
 \end{align}

\vspace{.4cm}
\noindent{\bf  Finite sizes.}
The calculations of the stability exponents,  Eqs. (\ref{exps}), apply only in the {\it asymptotic limit}, for infinite samples and  very close to the multicritical point, i.e., at very large $\ell$.
The explanation of the experiments (carried out at a finite $\xi$) and simulations (accomplished at a finite $L$) requires  the usage of a {\it finite number} of RG iterations,
 $\ell=\ell^{}_f$, at which the fluctuations have been eliminated: The renormalized correlation length $\xi(\ell^{}_f)={\cal O}(1)$,
 with $\xi(0)\sim|t|^{-\nu}$ ($t=T/T^{}_c-1$ measures the distance from the transition temperature $T_{c}$,  and  $\nu\approx .711$ is the critical exponent),
or the system size $L(\ell^{}_f)={\cal O}(1)$~\cite{RG}
 (lengths are measured in units of the lattice constant).
$\ell^{}_f$ increases  with the system's size $L$ (at criticality), or when the initial  parameters are  closer to criticality (i.e., a larger initial correlation length). At this stage, one can solve the problem using the mean-field Landau theory~\cite{RG}.
An  analysis of this situation requires the {\it full RG flow} of the system's Hamiltonian~\cite{referee}.  Such an analysis, based on resummation of (approximate) second-order $\epsilon-$expansions, was performed by Folk {\it et al.}~\cite{folk}.
That paper presented numerical RG flows in the parameter space, and observed the slow flow  close to  the isotropic and biconical  FP's.

\begin{figure}[htb]
\vspace{-2cm}
\includegraphics[width=0.4\textwidth]{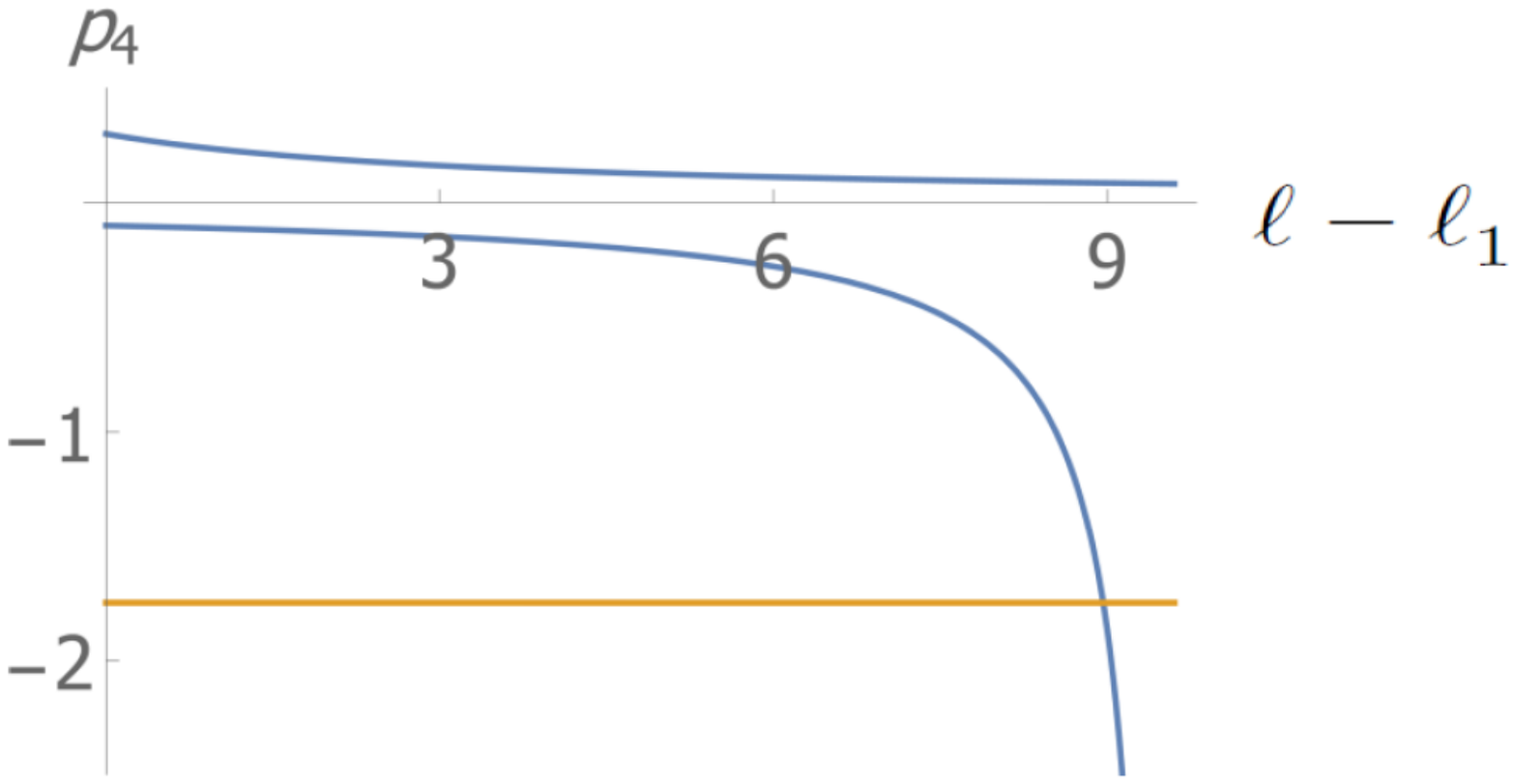}
\vspace{-5.5cm}
\caption{(color online)  The function $p^{}_4(\ell-\ell^{}_1)$  (blue) for $B=1$ and $p^{}_4(\ell^{}_1)=.3$ and $-.1$. Below the horizontal (orange) line  at $p^{}_4=-35 u^I/8=-1.75$, the transition becomes first order and the bicritical point becomes a triple point. }
\label{2}
\end{figure}

\vspace{.4cm}
\noindent{\bf  Our calculation.} This Letter  presents a more precise way to perform this analysis, based on  the following steps. (1) Using the stability exponents of the isotropic FP {\it at} three dimensions, Eq. (\ref{exps}), we construct  flow recursion relations {\it near} that FP. (2)  Equating  Eq. (\ref{U4}) with Eq. (\ref{U4n}), 
  the initial quartic parameters $\{u^{}_i\}$ are expressed in terms of the $p^{}_i$'s, with coefficients true to {\it all} orders in $\epsilon$ [see Eq. (\ref{uuu})  below].
 (3) Since $p^{}_0$ and $p^{}_2$ are strongly irrelevant ($\lambda^{}_0$ and $\lambda^{}_2$  are  negative and large [Eq. (\ref{exps})]) near the isotropic FP, they decay after a small number $\ell^{}_1$ of `transient' RG iterations (irrespective of non-linear terms in their recursion relations). After that,  the RG iterations continue on a {\it single universal} straight line in the three-dimensional parameter space,  given in Eq. (\ref{line}).
 In  a way, this line generalizes  the concept of universality. (4) On this universal line, Eq. (\ref{pil})  for $p^{}_{4}$ yields a slow flow  [as $p^{}_4(\ell)\sim e^{\lambda^{}_4\ell}$]  away from the isotropic FP for both positive and negative $p^{}_4$.
  The smallness of $\lambda^{}_{4}$ allows us to expand in powers of $p^{}_4$ around the isotropic FP [instead of the `usual' expansion in all the $u$'s near the Gaussian FP]. To second order in $p^{}_4$ [for $\ell>\ell^{}_1$],
 \begin{align}
 dp^{}_4/d\ell=\lambda^{}_4p^{}_4-Bp^2_4,
 \label{rrp4}
 \end{align}
 where the (positive) coefficient $B$ (the only unknown parameter) is presumably of order $1$. This yields explicit solutions for $p^{}_4(\ell)$, Eq. (\ref{flow}), and typical solutions are shown in Fig. \ref{2}.
 (5) For $p^{}_4>0$ the trajectories flow to the stable biconical FP, and the stability exponents at that point agree (approximately) with the full calculation in Ref. \onlinecite{vicari} -- adding credibility to our approximate expansion. On these trajectories the coefficients are shown to yield a tetracritical phase diagram.
 (6) For $p^{}_4<0$ the trajectories eventually flow to a fluctuation-driven first-order transition, which occurs when $p^{}_4(\ell)$ crosses the horizontal line in Fig. \ref{2}. In the wide intermediate range of $\ell$, before that crossing, the parameters yield a bicritical phase diagram.
 Beyond that crossing, for very large $\ell$ (corresponding to very large $L$ or $\xi$) the bicritical point turns into a triple point.  The bicritical phase-diagrams observed in the experiments/simulations  apparently occur  at this intermediate range.

\vspace{.4cm}
\noindent{\bf  Criteria for tetracriticality.}
Eliminating the small (non-critical) paramagnetic moment (generated by $H^{}_\parallel$) from the free energy renormalizes the three $u$'s in Eq. (\ref{U4}), with corrections of order $H^2_\parallel$~\cite{KNF}.  Although these corrections are small, so that the new coefficients remain close to the isotropic $u$, they are important because they determine the ultimate shape of the phase diagram. The tetracritical phase diagram [Fig. \ref{1}(b)] requires that on the line $g=0$ both order parameters are non-zero, implying that the mean-field free energy has a minimum at $0<x<1$~\cite{comg}.  Presenting Eq. (\ref{U4}) as
\begin{align}
U^{}_4=|{\bf S}|^4\big[ u^{}_\parallel x^2+ u^{}_\perp (1-x)^2+2 u^{}_\times x(1-x)\big]\ ,
\label{U4a}
\end{align}
 this minimum is at
 $x=( u^{}_\perp- u^{}_\times)/( u^{}_\parallel+ u^{}_\perp-2u^{}_\times)$,  provided that
\begin{align}
u^{}_\times< u^{}_\parallel\ \  {\rm and}\ \  u^{}_\times<u^{}_\perp.
\label{newcrit}
\end{align}
 These conditions for tetracriticality are more restrictive than the condition found before, $u^{}_\parallel u^{}_\perp-u^2_\times>0$~\cite{KNF}.  When even one of them is violated,  the minimum  of $U^{}_4$ is at $x=1$ or at $x=0$, implying  that the mixed phase does not exist; it is replaced by a first-order transition line, as in  Figs. \ref{1}(a,c).

\vspace{.4cm}
\noindent{\bf Renormalization group.}
  Comparing Eqs. (\ref{U4})  and (\ref{U4n})   for $U^{}_4$ one finds
\begin{align}
&\delta u^{}_\parallel=p^{}_0 + (70 p^{}_2 + 24 p^{}_4)/105,\nonumber\\
&\delta u^{}_\perp= p^{}_0 - (35 p^{}_2 - 9 p^{}_4)/105,\nonumber\\
&\delta u^{}_\times=p^{}_0 + (35 p^{}_2 - 72 p^{}_4)/210,
\label{uuu}
\end{align}
with $\delta u^{}_i=u^{}_i-u^I$.
According to Eq. (\ref{newcrit}), the multicritical point is tetracritical if both anisotropy parameters $u^{}_\parallel-u^{}_\times=p^{}_2/2+4p^{}_4/7$ and $u^{}_\perp-u^{}_\times=-p^{}_2/2+3p^{}_4/7$   are positive, i.e., when  $|p^{}_2(\ell)|<6p^{}_4(\ell)/7$.
Since $p^{}_2(\ell)\approx p^{}_2(0)e^{\lambda^{}_2\ell}$ decays rather quickly, and $p^{}_4(\ell)$ varies slowly (see below), this will happen when  $e^{\lambda^{}_2\ell}<6p^{}_4(0)/[7|p^{}_2(0)]$.
Assuming that  $p^{}_4(0)[=u^{}_\parallel+u^{}_\perp-2u^{}_\times]$ and $p^{}_2(0)[=2(3u^{}_\parallel-4u^{}_\perp+u^{}_\times)/7]$
are small and of the same order, this happens for a small $\ell< \ell^{}_1$. We conclude that the phase diagram is  in fact tetracritical whenever $p^{}_4(0)>0$, for practically all $\ell$, irrespective of the value of $B$. Since the experiments and simulations do not exhibit this phase diagram, we conclude  that they  probably have $p^{}_4(0)<0$.

To complete the RG analysis, we note that both
 $p^{}_0$ and $p^{}_2$ decay quickly, so there is no need to add higher-order terms for them in  Eq. (\ref{pil}).  They can be neglected in Eq. (\ref{uuu})  after a transient stage of  $\ell^{}_1$ iterations~\cite{el1},  and  then all  the flows continue on the universal semi-asymptotic line,
\begin{align}
\big(\delta u^{}_\parallel,~\delta u^{}_\perp,~\delta u^{}_\times\big)= \big(8,~ 3,~  -12 \big)p^{}_4/35.
\label{line}
\end{align}
 Higher-order terms in the RG recursion relations may turn this line  non-linear~\cite{AEK}.

For $\ell>\ell^{}_1$  the recursion relation for $p^{}_4$, Eq. (\ref{rrp4}),  gives the solution~\cite{AEK}
\begin{align}
p^{}_4(\ell)=\frac{p^{}_4(\ell^{}_1)e^{\lambda^{}_4(\ell-\ell^{}_1)}}{1+Bp^{}_4(\ell^{}_1)(e^{\lambda^{}_4(\ell-\ell^{}_1)}-1)/\lambda^{}_4}.
\label{flow}
\end{align}
For $p^{}_4(\ell^{}_1)>0$, the flow approaches the biconical FP, $p^{}_4(\ell)\rightarrow p^{B}_4=\lambda^{}_4/B$, with $p^B_4\ll 1$ -- justifying stopping the expansion in Eq. (\ref{rrp4}) at second order~\cite{BBB,commcub}.
  Near the biconical FP  one finds that (to linear order in $p^{}_4-p^B_4$) $d[p^{}_4-p^B_4]/d\ell=-\lambda^{}_4[p^{}_4-p^B_4]$, identifying the stability exponent at this FP as $\lambda^B_4=-\lambda^{}_4\approx-0.01$, independent of $B$, and the biconical FP is indeed stable.
  Within our approximate recursion relations for $p^{}_0$ and $p^{}_2$, the other two exponents approximately remain unchanged, $\lambda^B_{0,2}\approx\lambda^{}_{0,2}$. All three values are  close to those found near the biconical FP by the full sixth-order calculation in  Ref. \onlinecite{vicari}, confirming the validity of our approximate expansion near the isotropic FP.

 For $p^{}_4(\ell^{}_1)<0$, Eq. (\ref{rrp4}) implies that $p^{}_4(\ell)$ grows more and more negative (note: both $B$ and $\lambda^{}_4$ were assumed to be positive). At $\ell=\ell^{}_f$, Eq. (\ref{newcrit}) is not obeyed, the minimum of $U^{}_4$ is at $x=1$, with  $U^{}_{4,min}=|{\bf S}|^4u^{}_\parallel=|{\bf S}|^4[u^I+8p^{}_4(\ell^{}_f)/35]$, where we used Eq. (\ref{line}). This  becomes negative when $p^{}_4(\ell^{}_f)<-35u^I/8$. The resummation of the $\epsilon-$expansion gives $u^I\sim 0.4$~\cite{AEK}, leading to $35 u^I/8\sim 1.75$ [the orange horizontal line in Fig. \ref{2}], which is quite large compared to reasonable values of $p^{}_4(\ell^{}_1)$, and probably out of the region of applicability of the quadratic approximation which yielded  Eq. (\ref{flow}). However, it may still be reasonable for intermediate values of $\ell$ (e.g., $\ell-\ell^{}_1<8$ in Fig. \ref{2}).
Equation (\ref{flow}) diverges at a large $\ell=\ell^{}_2$~\cite{comm},
and we expect $p^{}_4(\ell)$ to cross the value $-1.75$ not very far below $\ell^{}_2$.  With the parameters used in Fig. \ref{2}, the divergence occurs at $\ell^{}_2-\ell^{}_1\sim\log[1-\lambda^{}_4/(Bp^{}_4(\ell^{}_1)]/\lambda^{}_4\sim 9.5$, and the transition to first-order occurs at $\ell^{}_x-\ell^{}_1\sim 9$. These numbers become smaller for larger values of $Bp^{}_4(\ell^{}_1)$.
In this example, the bicritical point turns into a triple point at $\xi\sim e^{\ell^{}_x}\sim e^{8+9}\sim 10^7$, which  cannot be reached experimentally.
Even if this approximation is improved, and if $B p^{}_4(0)$ increases (see the end of the paper), there will still be a wide range of parameters where experiments and simulations will follow the bicritical phase-diagram.
In this range,  the effective exponents near the bicritical point may depend on $\ell^{}_f$ and differ significantly from their  isotropic-FP values~\cite{AEK}.

\vspace{.4cm}
 \noindent{\bf Other examples.} Similar phase diagrams pertain to the structural transitions in uniaxially stressed perovskites, which are described by the cubic model~\cite{AEK,AAc,bruce}.
 Similarly to the XXZ antiferromagnet,  the almost isotropic SrTiO$^{}_3$ (with $p^{}_4\lessapprox 0$) yielded an apparent  bicritical phase diagram. However,  the more anisotropic  RbCaF$^{}_3$ did yield the diagram \ref{1}(c), as expected by the RG calculations~\cite{AEK}.

In  reality,  cubic anisotropic antiferromagnets are subjected to both the anisotropic  and cubic terms,  $U^{}_4$ and $U^{}_c$ (or other crystal-field terms). In most magnetic cases, the cubic terms are small~\cite{Shapira}. Since both ${\cal P}^{}_{4,4}$ and $U^{}_c$  scale with the same small exponent $\lambda^{}_4$, we expect the same qualitative flow diagrams as discussed above. However, the competition (within this subgroup) between the biconical and the cubic FP's (which are degenerate at linear order),    can only be settled  by  including higher-order terms in the RG recursion relations,  still awaits further analysis. Studies with other crystal symmetries (e.g., tetragonal), and detailed studies of the  sixth-order terms which dominate the fluctuation-driven tricritical point, also await a detailed analysis (and corresponding dedicated experiments).

For larger values of $n=n^{}_1+n^{}_2>3$, the biconical FP becomes unstable, being replaced by the {\it decoupled FP}, at which $u^D_\times=0$~\cite{AAD}, implying a tetracritical phase diagram. This has been particularly expected for the SO(5) theory aimed to describe the competition between superconductivity ($n^{}_1=2$) and antiferromagnetism ($n^{}_2=3$) in the cuprates~\cite{zhang}. In contrast,  Monte Carlo simulations  of this model gave a bicritical phase diagram, with isotropic $n=5$ critical exponents~\cite{hu}. Similar results were reported for the iron pnictides~\cite{pnic}. Assuming that the parameters of these materials  obey $u^{}_\times(\ell^{}_f)\nless u^{}_\parallel(\ell^{}_f),~u^{}_\perp(\ell^{}_f)$, preferring the bicritical scenario, and that the RG trajectories stay close to the isotropic FP, could also resolve that long-standing puzzle. 

A very recent experiment \cite{meta} studied a critical pressure-temperature phase diagram, with competing ferromagnetic and antiferromagnetic phases,  is also apparently in contrast to the RG results, which predict for $n^{}_1=n^{}_2=3$ an asymptotic decoupled tetracritical phase diagram (or a triple point). It would be interesting to study the RG trajectories for these experiments.

Competing order parameters, with larger values of $n$,  also arise in certain field-theory models~\cite{boot,QFT}, which are similar
in structure to the standard
model of particle interactions. It would be interesting to see whether those theories yield puzzles of the sort discussed here.

\vspace{.4cm}
\noindent{\bf Summary.}  In conclusion, experiments and  simulations do not contradict the renormalization-group predictions.   The  new system of recursion relations  presented here,  which is based on  group-theoretical exact coefficients  for an expansion near the isotropic fixed point, clearly  shows that  the simulations and experiments are in a crossover regime, between the bicritical point and the triple point. Our quantitative estimates show that it will probably be very difficult to reach  the triple point experimentally.  However, in principle the renormalization-group also supplies intermediate effective exponents~\cite{AEK}, whose measurements can confirm its validity. Dedicated experiments (carried out on larger samples, at temperatures closer to the multicritical point), and exploiting a wider range of  the initial Hamiltonians, which will allow increasing $p^{}_4(0)$ by moving away from the  parameters characterizing the isotropic fixed point (e.g., by adding single-ion anisotropies~\cite{selke1}), may find the tetracritical or the triple point, or -- at least -- detect the variation of the non-asymptotic (effective) critical exponents.

\vspace{.4cm} 
{\bf Acknowledgement:} We thank Andrey Kudlis, Walter Selke,  David Landau
 and Andrea Pelissetto for helpful correspondence.


\begin{thebibliography}{99}
\bibitem{comg} To identify the mixed phase it is sufficient to find it for $g=0$.  Keeping that term allows  to calculate the boundaries of the mixed phase in Fig. \ref{1}(b), and  is  important for calculating the crossover exponents which determine the shapes of the lines in the phase diagrams in Fig. \ref{1}~\cite{KNF,bruce}. These tasks are beyond the scope of the present paper.
\bibitem{el1} Assuming that these coefficients can be neglected when their magnitude is smaller than 1/1000, we end up with $\ell^{}_1\sim\max[\ln(1000p^{}_{2,4}(0))/\lambda^{}_{2,4}]$. This yields $\ell^{}_1\sim 8$ if $p^{}_0$ and $p^{}_2$ are of order 0.1. These transients then disappear for $\xi$ or $L$ of order $e^{\ell^{}_1}\sim e^8\sim 3000$.
\bibitem{BBB} The coefficient $B$ can  be deduced from $p^B_4$. Unfortunately, present papers, e.g., Ref. \onlinecite{vicari}, report only the universal exponents, and not the values of the FP parameters. $B$ can also be deduced from a resummation of the quadratic coefficients in the recursion relations near the isotropic FP~\cite{AEK}.
\bibitem{commcub} Reference \onlinecite{AEK} performed a similar analysis for the  cubic Hamiltonian, which  contains a single anisotropic term, $v\sum_{m=1}^3(S^{}_m)^4$. A numerical resummation of the linear scaling fields from the sixth-order $\epsilon-$expansion yielded the semi-asymptotic universal line $u=-0.595v$, while group-theory based arguments give $u=-3v/5$,  corroborating
 the resummation techniques. That calculation also resummed the quadratic terms in the recursion relations, yielding the analog of the coefficient $B$ here. 
\bibitem{comm} The expression in Eq. (\ref{flow}) diverges at  $\ell^{}_2$, when $1/p^{}_4(\ell^{}_1)+(e^{\lambda^{}_4(\ell^{}_2-\ell^{}_1)}-1)/p_4^B=0$.   Therefore, this approximation may not apply for very large $\ell^{}_f$. However, the qualitative behavior is expected to remain valid after adding higher orders in the flow equations. It certainly remains true if $\ell^{}_f  \ll \ell^{}_2$. Assuming Eq. (\ref{flow}), $B=1$ and $p^{}_4(\ell^{}_1)=.1$ the divergence happens at $\ell^{}_2-\ell^{}_1\sim 10$. This value decreases for larger values of $B p^{}_4(\ell^{}_1)$.
\end{thebibliography}
\end{document}